\documentclass[journal]{IAENGtran}
\usepackage{cite}

\ifCLASSINFOpdf
   \usepackage[pdftex]{graphicx}
   \DeclareGraphicsExtensions{.pdf,.jpeg,.png}
\else
   \usepackage[dvips]{graphicx}
   
   \DeclareGraphicsExtensions{.eps}
\fi
\usepackage{amsmath}
\usepackage{amssymb}
\usepackage{bm}

\begin{document}
%
\title{Determining the Number of Sinusoids
	Measured with Errors}
%
%
%

\author{Aleksandr~Kharin,~Oleg~V.~Chernoyarov
\thanks{
Manuscript received March 27, 2019; revised June 12, 2019. This work was financially supported by the Ministry of Education and Science of the Russian Federation (research project No. 2.3208.2017/4.6).}
\thanks{A. Kharin is with the Target Search Lab of Groundbreaking Radio Communication Technologies of
Advanced Research Foundation, Voronezh, Russia (corresponding author, phone: +79102888124; e-mail: aleksandr.kharin.1989@gmail.com; ORCID: orcid.org/0000-0002-4534-7046).} 
\thanks{O. V. Chernoyarov is with the Department of Electronics and Nanoelectronics, National Research University “MPEI”, Moscow, Russia as well as with the International Laboratory of Statistics of Stochastic Processes and Quantitative Finance, National Research Tomsk State University, Tomsk, Russia. He is also with the Department of Higher Mathematics and System Analysis, Maikop State Technological University, Maikop, Russia (e-mail: chernoyarovov@mpei.ru).}}

\maketitle

\pagestyle{empty}
\thispagestyle{empty}

\begin{abstract}
This paper describes how a priori information about the signal parameters can influence the accuracy of estimating the number of these signals. This study considers sinusoidal signals and it is supposed that the parameters (amplitudes, frequencies and phases) of the received signals are known up to a certain error. The error probability of the maximum likelihood estimation of the number of sinusoids is calculated under this condition.
\end{abstract}

\begin{IAENGkeywords}
model selection, sinusoids in noise, number of signals estimating, maximum likelihood, abridged error probability, quasi-likelihood method.
\end{IAENGkeywords}

%
\IAENGpeerreviewmaketitle

\section{Introduction}\label{sec1}
\IAENGPARstart{I}n this paper, the problem is studied of estimating the number of sinusoidal (modulated sinusoidal in general case) signals with unknown parameters. It is a fundamental problem in signal processing and its related areas. There are a lot of classical \cite{1AIC,2Schwarz1978,3Rissanen1978,4Wax1985} and modern \cite{5Stoica2004,6Nadler2011,7Ding2011,8Stoica2013,9Mariani2015,10TrifonovKharin1,12TrifonovKharin3,11TrifonovKharin2,13TrifonovKharin4,14TrifonovKharin5,15Badawy2017,16Ince2018} studies dealing with this problem. Most of them focus on the cases when the signal parameters are completely unknown. The problems that are described there cannot be solved by the classical maximum likelihood method, that is why many different approaches to modify the maximum likelihood method have been proposed. However, as it is shown in \cite{10TrifonovKharin1} and \cite{12TrifonovKharin3}, in some cases the maximum likelihood method can be applied directly without any modification. One of these cases is a problem of determining the number of signals with known parameters, but it is a very rare one in practice. Usually, there are cases when the signal parameters are measured with some errors. It means that the signal parameters are not completely unknown and we have some measured parameter values that can be used instead of the true ones in the likelihood function for the further estimating the number of signals.

Let us describe the steps for this approach implementation. Firstly, it is needed to synthesize the maximum likelihood (ML) algorithm for estimating the number of signals under the assumption that all the signal parameters are known. Next, one has to substitute some arbitrary parameter values in this algorithm for the true ones. For example, one can substitute the results of the measurement of these parameters. The resulting algorithm is called the quasi-likelihood (QL) one \cite{14TrifonovKharin5} for estimating the number of signals. In other words, the structure of the QL algorithm coincides with the structure of the ML algorithm for estimating the number of signals with the known parameters, except the values of the signal parameters. In the ML algorithm, the true values are used, while in the QL algorithm one uses some arbitrary values that may deviate from the true values.

The estimation of the signal parameters is not prerequisite for the QL algorithm functioning. This fact implies computational simplicity and other advantages of the QL algorithm that are described in this paper. The main disadvantage of the QL algorithm is the dependence of the performance of this algorithm on the value of the deviation of the measured parameters from the true ones. One needs to understand this dependence before using the QL algorithm.         

This paper is focused on the design and the performance analysis of the QL algorithm for estimating the number of modulated sinusoidal signals. In other terms, there is provided the performance analysis of the maximum-likelihood algorithm for estimating the number of signals in case when the signal parameters deviate from their true values. Note that, in our previous research \cite{14TrifonovKharin5} we analyzed a special case when only the amplitudes and phases of modulated sinusoids may deviate from their true values. In this paper a general case is examined: we suppose that all the parameters of the modulated sinusoids may deviate from their true values and we also provide the simulations to confirm our theoretical results.

The main goal of this paper is to study the dependency of the performance of the QL algorithm on such key factors as the measurement errors when measuring the signal parameters, signal-to-noise ratios (SNR), etc. The results of this study will help one to decide whether it is necessary to substitute the maximum-likelihood algorithm by some other one or not.

The structure of the paper includes the following parts: the problem statement and the structure of the QL algorithm are introduced in Section II; the theoretical performance analysis along with the universal approximation of the error probability is presented in Section III; and finally, Section IV, there are described the simulations that both represent and confirm the theoretical results.

\section{Problem formulation}\label{sec2}
Let us start with describing a model of the signal under analysis. It is composed of a number of the modulated sinusoids as follows
\begin{equation} \label{SinModel} 
	s(t, \nu ,{{\mathbf{a}}_\nu }, {{\bm{\omega }}_\nu }, {{\bm{\varphi }}_\nu }) = \sum\limits_{i = 1}^{\rm{\nu}} {a_i f_i(t){\cos}\!\left( \omega_i t - \varphi_i + \Psi_i(t)\right) },
\end{equation} 
where ${\rm \nu } \in \{1,..,\nu_{\max}\}$ denotes a number of sinusoids; $a_i \in \mathbb{R}^1$ is the amplitude; $\omega _i \in \left[0,2\pi \right)$ 
is the frequency, and ${\varphi _i \in \left[0,2\pi \right)}$ is the phase shift of the $i$-th sinusoid, respectively; $f_i(t)$, $\Psi_i(t)$ denote the amplitude and phase envelope, respectively; while the vectors ${{\mathbf{a}}_\nu } = \left( {{a_i}} \right)_{i = 1}^\nu $, $ {{\bm{\omega }}_\nu } = \left( {{\omega _i}} \right)_{i = 1}^\nu$, ${{\bm{\varphi }}_\nu } = \left( {{\varphi _i}} \right)_{i = 1}^\nu$ contain all amplitudes, frequencies and phases of the $\nu$ modulated sinusoids. In a trivial case, i.e., when $f_i(t)=1$, $\Psi_i(t)=0$ for all $i$ and $t$, the model \eqref{SinModel} reduces to the sum of sinusoids.

Suppose that the observed data $x(t)$ be a signal \eqref{SinModel} corrupted by the additive white Gaussian noise. In this case, one can represent $x(t)$ as
\begin{equation} \label{SinNoiseModel} 
	x(t) = \sum\limits_{i = 1}^{\rm{\nu_0}} {a_{0i} f_i(t){\cos}\!\left( \omega_{0i} t - \varphi_{0i} + \Psi_i(t)\right) } + \sigma n(t),
\end{equation} 
where by the subscript $0$ we denote the true values of the relevant unknown parameters.

Next let us write down the log-likelihood function for the model described by \eqref{SinModel} and \eqref{SinNoiseModel}. By applying the classical results from \cite{17Anderson2003} one can obtain
\begin{equation} \label{L_Gen}
	\begin{aligned}
		L(\nu ,{\bf{a}}_\nu,{\bm{\omega }}_\nu, {\bm{\varphi }}_\nu)& = \frac{1}{\sigma^2}\sum\limits_{t = 1}^{N_s}{x(t)s(t, \nu ,{{\mathbf{a}}_\nu }, {{\bm{\omega }}_\nu }, {{\bm{\varphi }}_\nu })}\\&-\frac{1}{2\sigma^2}\sum\limits_{t = 1}^{N_s}{s^2(t, \nu ,{{\mathbf{a}}_\nu }, {{\bm{\omega }}_\nu }, {{\bm{\varphi }}_\nu })}.
	\end{aligned}
\end{equation} 

In this paper, it is assumed that amplitudes, frequencies, and phases of the signals in \eqref{SinModel} are measured with errors. Thus, there used some values $a_i^*=a_{0i}+\Delta_{ai}$, $\omega_i^*=\omega_{0i}+\Delta_{\omega i}$, $\varphi_i^*=\varphi_{0i}+\Delta_{\varphi i}$ instead of the true values of these parameters where $\Delta_{ai}$, $\Delta_{\omega i}$ and $\Delta_{\varphi i}$ characterize the absolute errors in the measurement of the respective parameters. Let us substitute \eqref{SinModel} into \eqref{L_Gen} with the measured values of the unknown parameters 

\begin{equation} \label{Lx}
	\begin{aligned}
		L(\nu &,{\bf{a}}_\nu ^*,{\bm{\omega }}_\nu ^*, {\bm{\varphi }}_\nu ^*) =\\=& \frac{1}{\sigma^2}\sum\limits_{i = 1}^\nu  {a_i^* \sum\limits_{t = 1}^{N_s} {x(t){f_i}(t)\cos ({\omega _i^*}t + {\Psi_i}(t) - \varphi _i^*) } }\\-& \frac{1}{2\sigma^2}\sum\limits_{i = 1}^\nu  {\sum\limits_{j = 1}^\nu  {a_i^*a_j^*} } K_{i,j}^{**},
	\end{aligned}
\end{equation} 
where 
\begin{equation} \label{Kzz1}
	K_{i,j}^{**}=\sum\limits_{t = 1}^{N_s}{\cos ({\omega _i^*}t + {\Psi_i}(t) - \varphi _i^*)\cos ({\omega _j^*}t + {\Psi_j}(t) - \varphi _j^*)}.
\end{equation}
Hereinafter we use a comma to separate double indices from each other.

Now the QL algorithm for estimating the number of signals can be represented as
\begin{equation} \label{Alg}
	\hat{\nu}=\arg\min\left(-L(\nu ,{\bf{a}}_\nu ^*,{\bm{\omega }}_\nu ^*, {\bm{\varphi }}_\nu ^*) \right). 
\end{equation}

It is obvious that computational complexity of QL algorithm \eqref{Alg} is lower than that of other commonly used algorithms. This advantage of QL approach follows from the fact that the QL algorithm \eqref{Alg} does not require the calculation of the unknown parameters. However, it is also obvious that the performance of the QL algorithm \eqref{Alg} is decreasing with increasing the error in measurement of the signal parameters. That is why, it is especially important to have an instrument for the performance analysis of the QL algorithm \eqref{Alg}. Further, such instrument will be presented and tested.       

\section{Theoretical performance analysis}\label{sec3}
In this section, a theoretical performance analysis of the algorithm \eqref{Alg} is provided. 

Firstly, let us substitute the structure of the observed data \eqref{SinNoiseModel} into the log-likelihood function \eqref{Lx}
\begin{equation} \label{Lsn}
	\begin{aligned}
		L(\nu ,{\bf{a}}_\nu ^*,{\bm{\omega }}_\nu ^*, {\bm{\varphi }}_\nu ^*) = \frac{1}{\sigma^2}\sum\limits_{i = 1}^\nu \sum\limits_{j = 1}^{\rm{\nu_0}} {{a_i^*}{a_{0j}}K_{i,j}^{*}}+\frac{1}{\sigma^2}\sum\limits_{i = 1}^\nu  {a_i^*\eta_i} \\- \frac{1}{2\sigma^2}\sum\limits_{i = 1}^\nu  {\sum\limits_{j = 1}^\nu  {a_i^*a_j^*} K_{i,j}^{**}},
	\end{aligned}
\end{equation} 
where
\begin{equation} \label{etoeta}
	\eta_i=\sigma\sum\limits_{t = 1}^{N_s} {n(t){f_i}(t)\cos ({\omega _i^*}t + {\Psi_i}(t) - \varphi _i^*) },
\end{equation}
\begin{equation} \label{Kzz2}
\begin{aligned}
K&_{i,j}^{*}=\\
&\sum\limits_{t = 1}^{N_s}{\cos ({\omega _i^*}t + {\Psi_i}(t) - \varphi _i^*)\cos ({\omega _{0j}}t + {\Psi_{j}}(t) - \varphi _{0j})}.
\end{aligned}
\end{equation}
Note that the coefficients $K_{i,j}^{**}$ \eqref{Kzz1} and $K_{i,j}^{*}$ \eqref{Kzz2} can be represented as
\begin{equation} \label{Kzz}
	\begin{aligned}
		K_{i,j}^{**} =& {V_{ci,j}^{**}}\cos (\varphi _i^* - \varphi _j^*) + {V_{si,j}^{**}}\sin (\varphi _i^* - \varphi _j^*)\\ + &{W_{ci,j}^{**}}\cos (\varphi _i^* + \varphi _j^*) + {W_{si,j}^{**}}\sin (\varphi _i^* + \varphi _j^*),\\
		K_{i,j}^{*} =& {V_{ci,j}^{*}}\cos (\varphi _i^* - \varphi _{0j}) + {V_{si,j}^{*}}\sin (\varphi _i^* - \varphi _{0j})\\ + &{W_{ci,j}^{*}}\cos (\varphi _i^* + \varphi _{0j}) + {W_{si,j}^{*}}\sin (\varphi _i^* + \varphi _{0j}),
	\end{aligned}
\end{equation} 

and
{\footnotesize
	\[ \left(\!\!\begin{array}{c}{V_{ci,j}^{**}}\\{V_{si,j}^{**}}\end{array}\!\!\right)\! =\!\frac{1}{2} \sum\limits_{t = 1}^{N_s} f_{i} (t)f_{j} (t) \! \left(\!\!\begin{array}{c}{\cos}\\{\sin}\end{array} \!\!\right)\! \left(\left(\omega _{i}^* \!-\!\omega _{j}^* \right)\!t\!+\!\Psi _{i} (t)\!-\!\Psi _{j} (t)\right)\!,\]
	\[ \left(\!\!\begin{array}{c}{W_{ci,j}^{**}}\\{W_{si,j}^{**}}\end{array}\!\!\right)\!\!=\!\frac{1}{2} \sum\limits_{t = 1}^{N_s} f_{i} (t)f_{j} (t) \!\left(\!\!\begin{array}{c}{\cos}\\{\sin}\end{array}\!\!\right)\! \left(\left(\omega _{i}^* \! + \! \omega _{j}^* \right)\!t \! +\! \Psi _{i} (t) \!+\! \Psi _{j} (t)\right)\!,\]
	\[ \left(\!\!\begin{array}{c}{V_{ci,j}^{*}}\\{V_{si,j}^{*}}\end{array}\!\!\right)\!=\!\frac{1}{2} \sum\limits_{t = 1}^{N_s} f_{i} (t)f_{j} (t) \!\left(\!\!\begin{array}{c}{\cos}\\{\sin}\end{array}\!\!\right)\! \left(\left(\omega _{i}^*\!-\!\omega _{0j} \right)\!t\!+\!\Psi _{i}(t)\!-\!\Psi _{j}(t)\right)\!,\]
	\[ \left(\!\!\begin{array}{c}{W_{ci,j}^{*}}\\{W_{si,j}^{*}}\end{array}\!\!\right)\!=\!\frac{1}{2} \sum\limits_{t = 1}^{N_s} f_{i} (t)f_{j}(t)\!\left(\!\!\begin{array}{c}{\cos}\\{\sin}\end{array}\!\!\right)\! \left(\left(\omega _{i}^*\!+\!\omega _{0j}\right)\!t+\!\Psi _{i}(t)\!+\!\Psi _{j}(t)\right)\!.\]
	\par}
The representations \eqref{Kzz} are useful for the further calculations of the coefficients $K_{i,j}^{**}$ \eqref{Kzz1} and $K_{i,j}^{*}$ \eqref{Kzz2}.

In this paper, we use an error probability, i.e., $\Pr(\hat{\nu} \neq \nu_0) $ as a measure of performance of the algorithm for estimating the number of signals and an abridged error probability \cite{10TrifonovKharin1,11TrifonovKharin2,12TrifonovKharin3,13TrifonovKharin4,14TrifonovKharin5} as a universal approximation to the error probability. Let us write down a definition of the abridged error probability that can be found in \cite{10TrifonovKharin1,11TrifonovKharin2,12TrifonovKharin3,13TrifonovKharin4,14TrifonovKharin5} 
\begin{equation} \label{pa1}
	p_{a}\!=\! 1 \!-\! \Pr\!\left(  L\!\left( {\nu _0}\right) \!-\! L\!\left( {\nu _0-1}\right)\!>\! 0,L\!\left( {\nu _0}\right)\! -\! L\!\left( {\nu _0+1}\right)\!>\! 0 \right)\!,
\end{equation}
where $L\left( {\nu}\right)$ is a log-likelihood function, i.e., in our case one can write as
\[L\left( {\nu}\right)=L(\nu ,{\bf{a}}_\nu ^*,{\bm{\omega }}_\nu ^*, {\bm{\varphi }}_\nu ^*).\]

Now let us calculate an abridged error probability for the algorithm \eqref{Alg}. For this purpose, the log-likelihood function \eqref{Lsn} should be substituted into the definition of the abridged error probability \eqref{pa1}. After all the transformations, \eqref{pa1} can be represented as follows 
\begin{equation} \label{pa2}
	p_{a}= 1 - \Pr\!\left(  \xi _{\nu _0} > - R,\xi _{{\nu _0} + 1} < Q \right),
\end{equation} 
where
{\footnotesize
	\begin{equation*} 
		\begin{split}
			R&=\!\frac{\left(\sum\limits_{j = 1}^{\nu_0}{a_{0j}K_{{\nu_0}, j}^*} - \frac{1}{2}{a_{\nu_0}^*K_{{\nu_0},{\nu_0}}^{**}}-{\sum\limits_{j = 1}^{\nu_0-1}{a_j^*} K_{{\nu_0},j}^{**}}\right)}{{\sigma}\sqrt{K_{{\nu_0},{\nu_0}}^{**}}}, \\
			Q&=\!\frac{\left(\!-\!\sum\limits_{j = 1}^{\nu_0}{{a_{0j}}K_{{\nu_0+1}, j}^*} \!+\!{\frac{1}{2}}{a_{\nu_0+1}^*K_{{\nu_0+1},{\nu_0+1}}^{**}}\!+\!{\sum\limits_{j = 1}^{\nu_0}{a_j^*} K_{{\nu_0+1},j}^{**}}\!\right)}{{\sigma}\sqrt{K_{{\nu_0+1},{\nu_0+1}}^{**}}}, 
		\end{split}
	\end{equation*} \par}
and
\begin{equation*} 
	\xi_i = \eta_i/\left(\sigma \sqrt{K_{i,i}^{**}}\right). 
\end{equation*}

The following final formula for the abridged error probability of the algorithm \eqref{Alg} for estimating the number of modulated sinusoids can be obtained using some classical results from \eqref{pa2}. 
\begin{equation} \label{Final_pa}
	p_{\textrm{a}} = 1 - \frac{1}{{\sqrt {2\pi } }} \!\! \int\limits_{-\infty}^{Q} \!\! {\exp \! \left( { - \frac{{{y^2}}}{2}} \right)} \Phi\!\! \left(\frac{R + \rho y}{\sqrt{1 -\rho^2}}\right)\textrm{d}y,
\end{equation}
where
\begin{equation*}
	\rho=\frac{K_{{\nu_0},{\nu_0+1}}^{**}}{\sigma^2\sqrt{K_{{\nu_0},{\nu_0}}^{**}K_{{\nu_0+1},{\nu_0+1}}^{**}}},
\end{equation*}
and  
\begin{equation*}
	\Phi (x)=\frac{1}{\sqrt{2\pi } }\!\! \int \limits _{-\infty }^{x}\!\!\exp\! \left(-t^{2} /2\right)\textrm{d}t. 
\end{equation*}

Some practical applications may require reducing the computational complexity of the result \eqref{Final_pa}. An asymptotically exact approximation for \eqref{Final_pa} can be written as follow
\begin{equation}\label{approc_final_pa}
	\begin{split}
	{p_{\textrm{a}}} \simeq 1 &- \Phi \! \left( R \right) \! \Phi \! \left(Q\right)\\
	 &+ \frac{\rho}{2\pi} \exp \!  \left( { - \frac{R^2}{2}} \right) \! \exp  \! \left( { - \frac{Q^2}{2}} \right)\!.
	\end{split}
\end{equation}
The approximate formula \eqref{approc_final_pa} can be used instead of the exact formula \eqref{Final_pa} for the next parameters ranges

$$\min \! \left( Q, R \right)>3, \text{ and } \left| \rho \right|<0.9  .$$

Now let us describe how the result \eqref{Final_pa} can help to make the choice between the QL algorithm and any other algorithms. Suppose that the true values of the unknown parameters belong to some known intervals, i.e., for all $i$: $a_{0i} \in [a_{li}, a_{ri}]$, $\varphi_{0i} \in [\varphi_{li}, \varphi_{ri}]$ and $\omega_{0i} \in [\omega_{li}, \omega_{ri}]$. This is a typical case for measurements with errors. The result \eqref{Final_pa} can help one to choose between the quasi-likelihood algorithm and any other algorithms for estimating the number of signals, the following should be made:

1. Consider \eqref{Final_pa} as a function of $\Delta_{ai}$, $\Delta_{\omega i}$ and $\Delta_{\varphi i}$, i.e., $p_{\textrm{a}}(\Delta_{ai},\Delta_{\omega i},\Delta_{\varphi i})$.

2. Find the maximum of the $p_{\textrm{a}}(\Delta_{ai},\Delta_{\omega i},\Delta_{\varphi i})$ using $a_{li}, a_{ri}, \varphi_{li}, \varphi_{ri}, \omega_{li}, \omega_{ri}$ as the constraints. 

3. Find the abridged error probability for the algorithms that pretend to be alternatives to the QL algorithm.

4. By means of the results produced at the steps 1. -3. and having estimated the computational complexity of the tested algorithms, and some other arguments, one can make the choice.

\section{Abridged error probability}\label{sec4}
Next let us list the some basic properties of the abridged error probability that follow from its definition \eqref{pa1}. 

P.1) The abridged error probability is a lower bound of the error probability in the case $ 1<\nu_0<\nu_{max}$.

P.2) The abridged error probability is equal to the error probability in the case $ \nu_0=2$, $\nu_{max}=3$.  

 Assume that the absolute values of ${L({\nu _0} - 1)}$ and ${L({\nu _0} + 1)}$ increasing with the SNR or the number of samples increasing faster than all the terms from the set $\left\lbrace \left\lbrace {L(i)} \right\rbrace_{i=1}^{\nu_0 - 2} \bigcup \left\lbrace {L(i)} \right\rbrace_{i=\nu_0+2}^{\nu_{max}}\right\rbrace $, i. e., if the SNR or the number of samples trends to infinity, then 
\begin{equation} \label{assum_1} 
\begin{split}
\frac{\left| L(i)\right| }{\left| L( {\nu _0}-1)\right| } \rightarrow 0 \text{\,\,\,and\,\,\,}
\frac{\left| L(i)\right| }{\left| L( {\nu _0}+1)\right| } \rightarrow 0,
\end{split}
\end{equation}
for any $i \in \left\lbrace 1, \ldots, \nu_0 - 2, \nu_0 + 2, \ldots \nu_{max}\right\rbrace  $.

P.3) If the assumption \eqref{assum_1} is fulfilled, then the abridged error probability trends to error probability as SNR or number of samples tends to infinity.

Let us discuss the assumption \eqref{assum_1}. If this assumption is not fulfilled, then also the following condition does not hold:
\begin{equation} \label{cond_1} 
\frac{p\left( \left| \hat{\nu}-\nu_0 \right| >1 \right) }{p\left( \left| \hat{\nu}-\nu_0 \right| =1 \right)}\rightarrow 0,
\end{equation}
as the SNR or the number of samples trends to infinity.
  
The condition \eqref{cond_1} is important because the error situation when $ \left| \hat{\nu}-\nu_0 \right| =1$ is usually much more better on practice than the error situation when $ \left| \hat{\nu}-\nu_0 \right| >1$. Thus the assumption \eqref{assum_1} is usually fulfilled for the algorithms that is useful in practice.

\section{Simulations}\label{sec5}
Theoretical performance analysis of the QL algorithm \eqref{Alg} presented in Section III needs to be confirmed. In this section we describe the simulations of the algorithm \eqref{Alg}. These simulations provide an independent way to calculate the error probability for the algorithm \eqref{Alg}. 

To simplify the presentation of our simulations, it is presupposed that 

(i) For all $i$ and $t$, $f_i(t)=1$, $\Psi_i(t)=0$  (in this case the model \eqref{SinModel} reduces to the sum of sinusoids) 

(ii) For all $i$, $a_{0i}=a_{0}$, $\omega_{0i}=(i-1)\omega_{st}+\omega_{b}$, $\Delta_{ai}=\Delta_{a}$, $\Delta_{\omega i}=\Delta_{\omega}$ and $\Delta_{\varphi i}=\Delta_{\varphi}$.

In the provided simulations, the following true values of signal and noise parameters are used: $a_{0}=0.4$, $\omega_{st}=2\pi B/N_s$, $\omega_{b}=0.4\pi $, $\omega_{01}=1.2075$, $\omega_{02}=1.2566$, $\omega_{03}=1.3057$, $\omega_{04}=1.3548$, $\omega_{05}=1.4039$, $\varphi_{01}=0$, $\varphi_{02}=\pi/4$, $\varphi_{03}=\pi/3$, $\varphi_{04}=\pi/5$, $\varphi_{05}=\pi/6$.

For all the simulations we use the classical definition for the signal-to-noise ratio (SNR):
\[z=a_{0}^2/2\sigma, \,\,\, z\,\textrm{dB}=10\log_{10}\left(a_{0}^2/2\sigma\right). \]
There are also used the relative errors for describing both the amplitude error and the frequency error: $\delta_a=\Delta_{a}/{a_0}$ and $\delta_\omega=\Delta_{\omega}/\omega_{st}$, respectively.

In this paper, the simulations test the two cases:

(i) the true number of signals is $\nu_0=2$, while the maximum possible number of signals is $\nu_{\max}=3$. In this case, the abridged error probability \eqref{pa1} is equal by definition to the error probability. Therefore, one can check the theoretical formula for the  abridged error probability \eqref{Final_pa};

(ii) $\nu_0=3$, $\nu_{\max}=5$. This case is used to analyze the error probability approximation precision in terms of the abridged error probability.


The numerical studies are started with analyzing the approximation of the error probability in terms of the abridged error probability.  In this analysis we consider error probability as a function of the SNR. There has been performed $2\cdot10^5$ independent trials for each SNR. In Fig. 1. and 2, there are represented some typical examples of the results of these simulations for $N_s=128$ (Fig.1) and $N_s=512$ (Fig.2). Let us list the settings for Fig 1., Fig 2.: $\delta_a=0.25$, $\Delta_{\varphi}=0.1$, $\delta_\omega=0.02$. The following notations are used: the solid curve is the theoretical curve predicted by  \eqref{Final_pa}, the circles mark simulation results for $\nu_{\max}=3$ and $\nu_0=2$, and the squares mark simulation results for $\nu_{\max}=5$ and $\nu_0=3$.  

\begin{figure}[!t]
	\centering{\includegraphics{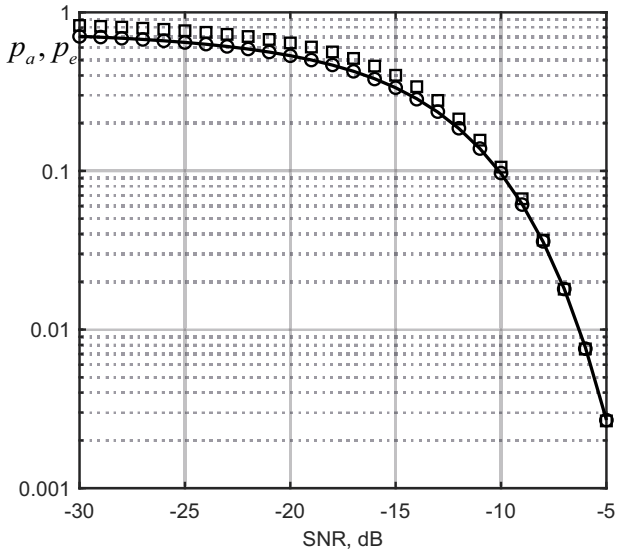}}
	\caption{Error probability in the number of signals estimation. Parameter settings: $N_s=128$, $\delta_a=0.25$, $\Delta_{\varphi}=0.1$, $\delta_\omega=0.02$. The following notations are used: the solid curve is the theoretical curve predicted by  \eqref{Final_pa}, the circles mark simulation results for $\nu_{\max}=3$ and $\nu_0=2$, and the squares mark simulation results for $\nu_{\max}=5$ and $\nu_0=3$. 
		\label{fig1}}
\end{figure}

\begin{figure}[!t]
	\centering{\includegraphics{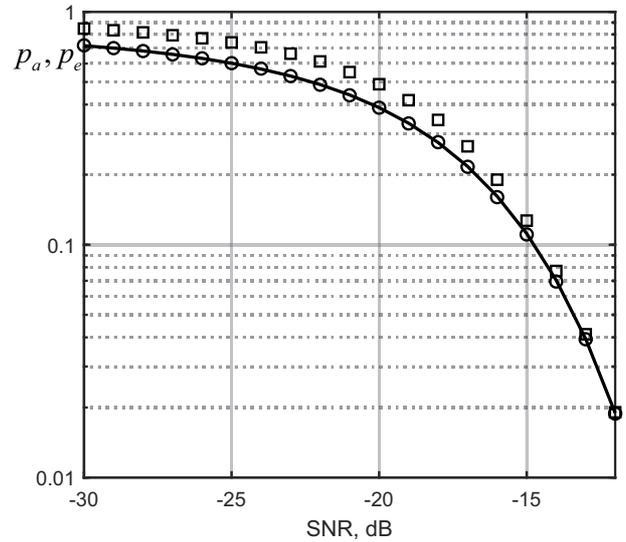}}
	\caption{Error probability in the number of signals estimation. Parameter settings: $N_s=512$, $\delta_a=0.25$, $\Delta_{\varphi}=0.1$, $\delta_\omega=0.02$. The following notations are used: the solid curve is the theoretical curve predicted by  \eqref{Final_pa}, the circles mark simulation results for $\nu_{\max}=3$ and $\nu_0=2$, and the squares mark simulation results for $\nu_{\max}=5$ and $\nu_0=3$.
		\label{fig2}}
\end{figure}

From Fig. 1 and Fig. 2 one can conclude, for example, that

(i)  the simulations presented in Fig. 1 and Fig. 2 confirm theoretical formula \eqref{Final_pa}; 

(ii) the error probability can be sufficiently approximated by the abridged error probability \eqref{Final_pa} starting with a particular SNR, for example: if $N_s=128$, $\nu_{\max}=5$ and $\nu_0=3$, then starting with $z=-11\textrm{ dB}$;

(iii) the error probability of the quasi-likelihood algorithm tends to zero with the SNR increasing (as well as when the number of samples is increasing), i.e., the algorithm \eqref{Alg} is a consistent estimator (for the above settings).    

The last conclusion can be obtained directly from the formula \eqref{Final_pa} that can also help to theoretically determine the errors in parameters so that the estimator \eqref{Alg} is a consistent one. It is assumed that the estimator is consistent if its error probability tends to zero with both the SNR and number of samples increasing. Only a few among the commonly known estimators satisfy both of these conditions \cite{7Ding2011}.  

Our further performance analysis is based on the theoretical formula \eqref{Final_pa} only. We set the SNR to $-11\textrm{ dB}$ for all the simulations that are presented in Fig. 3-7. 

In Fig. 3-5 we consider an abridged error probability \eqref{Final_pa} as a function of $\delta_a$, i.e., $p_{\textrm{a}}(\delta_a)$. Moreover in these and further figures we use the normalized abridged error probability that can be defined as $p_{\textrm{a}}(\delta_a)/p_{\textrm{a}}(0)$.

In Fig. 3. the case of $\Delta_{\varphi}=0$ is presented with the following designations for the different values of $\delta_\omega$: 
bold solid line for  $\delta_\omega=0$;
dash-dotted line for $\delta_\omega=0.08$;
dotted line for $\delta_\omega=0.12$;
dashed line for $\delta_\omega=0.16$;
solid line for $\delta_\omega=0.2$.

\begin{figure}[!t]
	\centering{\includegraphics{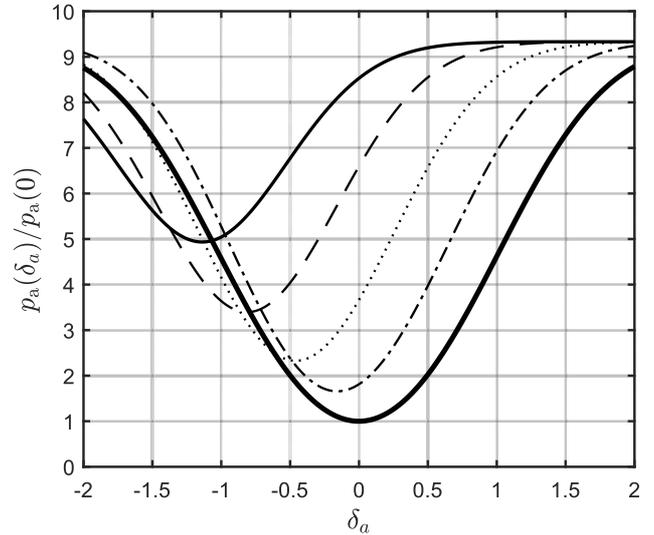}}
	\caption{Number of signals estimation accuracy losses due to the errors in the measurements of amplitudes and frequencies. Parameter settings: $N_s=128$, $\textrm{SNR}=-11\textrm{ dB}$, $\Delta_{\varphi}=0$. Line settings: bold solid line for  $\delta_\omega=0$; dash-dotted line for $\delta_\omega=0.08$; dotted line for $\delta_\omega=0.12$; dashed line for $\delta_\omega=0.16$; solid line for $\delta_\omega=0.2$.
		\label{fig3}}
\end{figure}
\begin{figure}[!t]
	\centering{\includegraphics{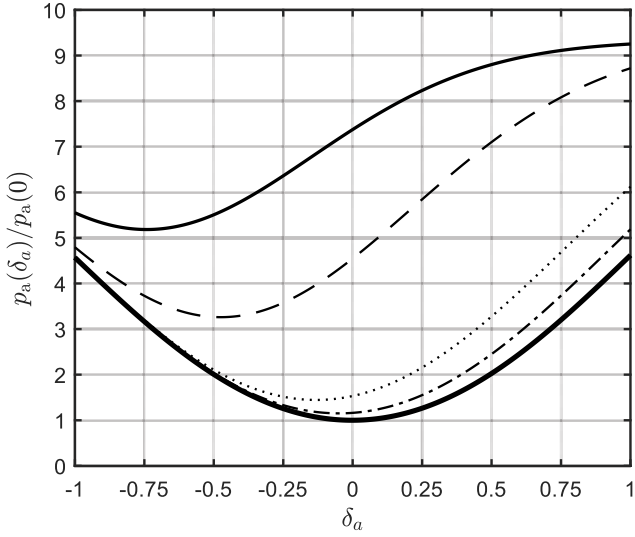}}
	\caption{Number of signals estimation accuracy losses due to the errors in the measurements of amplitudes and initial phases. Parameter settings: $N_s=128$, $\textrm{SNR}=-11\textrm{ dB}$, $\delta_\omega=0$. Line settings: bold solid line for $\Delta_{\varphi}=0$;
	dash-dotted line for $\Delta_{\varphi}=0.3$; dotted line for $\Delta_{\varphi}=0.5$; dashed line for $\Delta_{\varphi}=1$; solid line for $\Delta_{\varphi}=1.3$.
		\label{fig4}}
\end{figure}
\begin{figure}[!t]
	\centering{\includegraphics{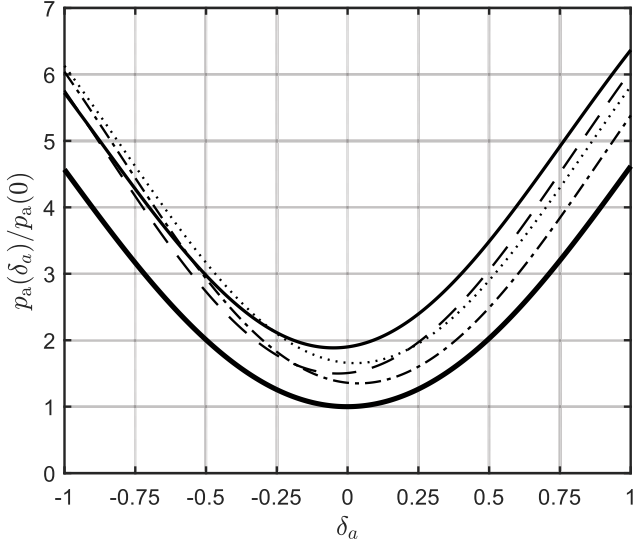}}
	\caption{Number of signals estimation accuracy losses due to the errors in the measurements of amplitudes, frequencies and initial phases. Parameter settings: $N_s=128$, $\textrm{SNR}=-11\textrm{ dB}$. Line settings: bold solid line  for $\Delta_{\varphi}=0$, $\delta_\omega=0$; dash-dotted line for $\Delta_{\varphi}=0.3$, $\delta_\omega=0.08$; dotted line for $\Delta_{\varphi}=0.5$, $\delta_\omega=0.12$; dashed line for $\Delta_{\varphi}=1$, $\delta_\omega=0.16$; solid line for $\Delta_{\varphi}=1.3$, $\delta_\omega=0.2$.
\label{fig5}}
\end{figure}
Fig. 4. shows the case when $\delta_\omega=0$. Here the following designations are used for the different values of $\Delta_{\varphi}$: 
bold solid line for $\Delta_{\varphi}=0$;
dash-dotted line for $\Delta_{\varphi}=0.3$;
dotted line for $\Delta_{\varphi}=0.5$;
dashed line for $\Delta_{\varphi}=1$;
solid line for $\Delta_{\varphi}=1.3$.

In Fig. 5. the mixed case is presented when all the parameter errors may have non-zero values, and the following designations are used for the  different values of $\Delta_{\varphi}$ and $\delta_\omega$:
bold solid line  for $\Delta_{\varphi}=0$, $\delta_\omega=0$;
dash-dotted line for $\Delta_{\varphi}=0.3$, $\delta_\omega=0.08$;
dotted line for $\Delta_{\varphi}=0.5$, $\delta_\omega=0.12$;
dashed line for $\Delta_{\varphi}=1$, $\delta_\omega=0.16$;
solid line for $\Delta_{\varphi}=1.3$, $\delta_\omega=0.2$.

Figs. 3-5 demonstrate that the case of the joint errors in three parameters can be rather than the case when the joint errors in two parameters take place. 

\begin{figure}[!t]
	\centering{\includegraphics{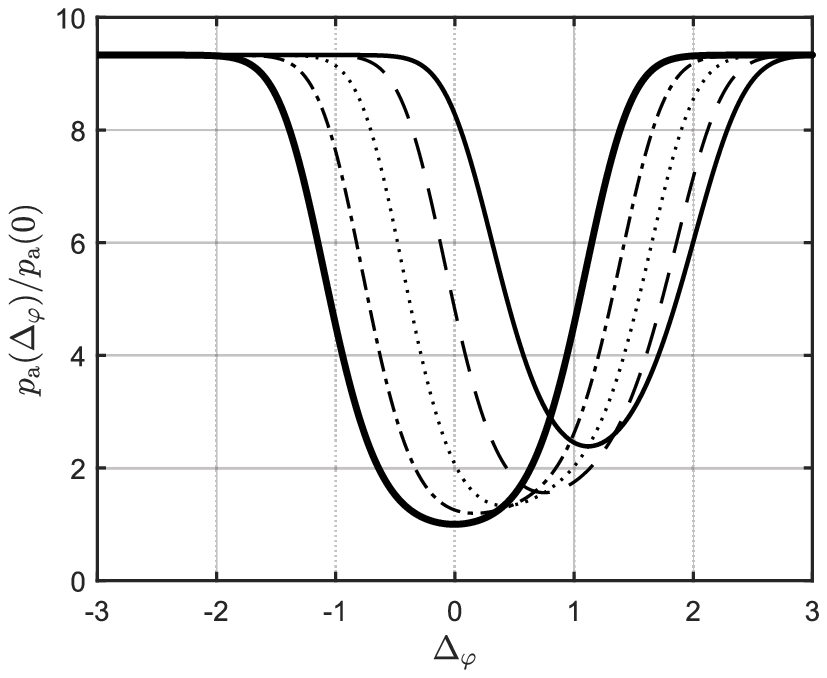}}
	\caption{Number of signals estimation accuracy losses due to the errors in the measurements of amplitudes, frequencies and initial phases. Parameter settings: $N_s=128$, $\textrm{SNR}=-11\textrm{ dB}$. Line settings: bold solid line  for $\delta_a=0$, $\delta_\omega=0$; dash-dotted line for $\delta_a=0.01$, $\delta_\omega=0.04$;	dotted line for $\delta_a=0.04$, $\delta_\omega=0.08$;	dashed line for $\delta_a=0.08$, $\delta_\omega=0.12$; solid line for $\delta_a=0.16$, $\delta_\omega=0.16$.
		\label{fig6}}
\end{figure}

\begin{figure}[!t]
	\centering{\includegraphics{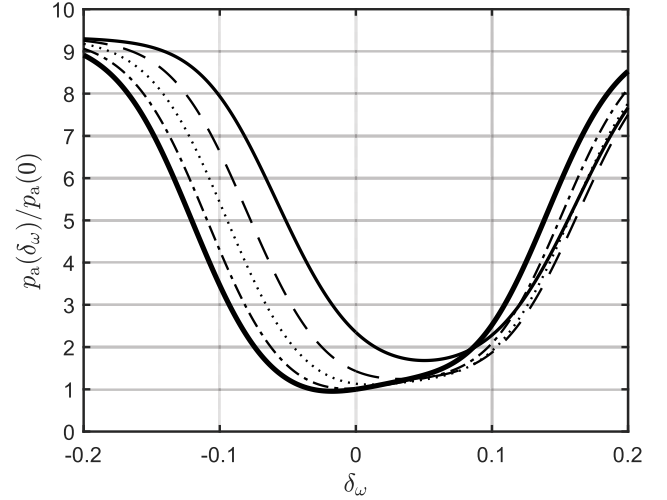}}
	\caption{Number of signals estimation accuracy losses due to the errors in the measurements of amplitudes, frequencies and initial phases. Parameter settings: $N_s=128$, $\textrm{SNR}=-11\textrm{ dB}$. Line settings: bold solid line for $\delta_a=0$, $\Delta_{\varphi}=0$;
	dash-dotted line for $\delta_a=0.01$, $\Delta_{\varphi}=0.1$; dotted line for $\delta_a=0.04$, $\Delta_{\varphi}=0.2$; dashed line for $\delta_a=0.08$, $\Delta_{\varphi}=0.3$; solid line for $\delta_a=0.16$, $\Delta_{\varphi}=0.4$.
		\label{fig7}}
\end{figure}

Figs. 6. and 7 also show mixed cases. In Fig. 6. the abridged error probability \eqref{Final_pa} is presented as a function of the error in measurement of the phase $\Delta_{\varphi}$, i.e., the normalized abridged error probability can be represented as   
\[p_{\textrm{a}}(\Delta_{\varphi})/p_{\textrm{a}}(0).\]
The following designations are used in Fig.6.:
bold solid line  for $\delta_a=0$, $\delta_\omega=0$;
dash-dotted line for $\delta_a=0.01$, $\delta_\omega=0.04$;
dotted line for $\delta_a=0.04$, $\delta_\omega=0.08$;
dashed line for $\delta_a=0.08$, $\delta_\omega=0.12$;
solid line for $\delta_a=0.16$, $\delta_\omega=0.16$.

Next, in Fig. 7. the abridged error probability \eqref{Final_pa} is represented as a function of the relative error in measurement the frequency $\delta_{\omega}$, and the normalized abridged error probability is rewritten as 
\[p_{\textrm{a}}(\delta_\omega)/p_{\textrm{a}}(0).\]
We use the following designations in. Fig.7.: 
bold solid line for $\delta_a=0$, $\Delta_{\varphi}=0$;
dash-dotted line for $\delta_a=0.01$, $\Delta_{\varphi}=0.1$;
dotted line for $\delta_a=0.04$, $\Delta_{\varphi}=0.2$;
dashed line for $\delta_a=0.08$, $\Delta_{\varphi}=0.3$;
solid line for $\delta_a=0.16$, $\Delta_{\varphi}=0.4$.

From Fig. 7 one can find that the errors in frequencies with the same values and the different signs may give different impacts on the abridge error probability of the algorithm \eqref{Alg}.

\section{Practical applications}\label{sec6}

Now let us write down a practical example. We suppose that the allowable normalized abridged error probability is less than $3$.  In this case, from Fig. 7. one can conclude that for all the handled cases of errors in the other parameters the relative error in measurement of the frequency should be less than $0.01$. Let also suppose that the sources of signals can move and that the error in frequency measurement is linked to the Doppler effect only. These conditions imply the following speed limit for our sources
\[v_s<0.00025c,\]
where $v_s$ is the speed of the source and $c$ is the wave speed.

If the emission from our sources is an electromagnetic emission, then $v_s<75000$ m/s. It is a very wide range. However, if the observed emission is a hydroacoustic emission, then the range of the allowable speed is very narrow one: $v_s<0.375$ m/s.

The provided example shows that the introduced QL algorithm \eqref{Alg} can be useful in many practical applications. Let us list some examples of these applications. Firstly, QL algorithm can be applied for the fast estimating of a multipath channel in the 5G \cite{Eluwole2018},  and other technologies (including IEEE 802.11p \cite{Shi2014}). Secondly, one may use QL algorithm for the fast DOA estimation in MIMO arrays \cite{DOA}. Next, the fast estimations of the parameters of the air targets \cite{Dong2016, Chen2018} can be obtained by the introduced QL algorithm. Note that, QL algorithm can be also useful for estimating the parameters of the underwater targets \cite{Lv2016} with the above speed limits. Finally, QL algorithm can be a part of the process monitoring algorithms \cite{Wang2019} and the system parameters estimation algorithms \cite{Wei2018}.

\section{Conclusion}\label{sec7}

In this paper, there is introduced the QL algorithm for estimating the number of signals with the unknown parameters. The QL algorithm can be obtained from the ML algorithm for estimating the number of signals with the known parameters by substituting some arbitrary values for the true values of the signal parameters. Next, there is presented both the theoretical and numerical performance analysis of the quasi-likelihood estimation of the number of signals. 
Firstly, a closed formula is theoretically obtained for the abridged error probability of the QL algorithm. 
Secondly, the obtained formula is confirmed by the simulations that are also used to provide a numerical study of the abridged error probability as an approximation to the error probability. 
Finally, the graphical analysis is presented of the obtained theoretical formula and some of its properties are found. The results that are obtained and confirmed in this paper can help in making a reasonable choice between the QL algorithm and any other algorithms for estimating the number of signals.


\bibliographystyle{BibTeXtran}   
\bibliography{BibTeXrefs}       

%

\begin{IAENGbiography}[{\includegraphics[width=1in,height=1.25in,clip,keepaspectratio]{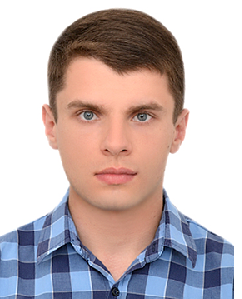}}]{Aleksandr Kharin}
was born in Voronezh, Russia in 1989. He received the B.S., M.S. and Phd degrees from the Department of Radiophysics, Voronezh State University, Voronezh, Russia, in 2010, 2012 and 2016, respectively, all in radiophysics. His main interests are statistical signal processing, estimation theory, applied probability and applied statistics. 

Dr. Kharin is IEEE Member since 2016.	
\end{IAENGbiography}







\end{document}